\documentclass{mema}
\usepackage{natbib}\usepackage{txfonts}
\usepackage{graphicx}
\usepackage[a4paper]{hyperref}
\idline{76/4}{1}
\begin{document}

\title{
Metallicity effects on distances based on double-mode Cepheids
}

   \subtitle{}

\author{
G.\, Kov\'acs\inst{} 
}

  \offprints{G. Kov\'acs}

\institute{
Konkoly Observatory, Budapest, Hungary \ 
\email{kovacs@konkoly.hu}
}

\authorrunning{Kov\'acs}

\titlerunning{Metallicity effects on distances}

%
%
\abstract{
We test the effect of systematic metallicity difference 
between the two types of double-mode Cepheids in the 
Magellanic Clouds. We show that lowering the metallicity 
for the short-periodic Cepheids leaves the distances at 
their canonical values and, at the same time, improves 
the agreement between the observed and model periods. 

}
\maketitle{}

%
%
\section{Introduction}

Double-mode variables are important in our understanding 
of the simplest form of steady multimode pulsations and they 
are also the prime targets in the simplest application of 
asteroseismology. In this latter aspect we investigated the 
use of the observed periods and color in deriving distances 
to various classes of variables (RR~Lyrae stars in globular 
clusters: Kov\'acs \& Walker 1999; RR~Lyrae stars in the 
Large Magellanic Cloud [LMC]: Kov\'acs 2000b; beat 
[i.e., double-mode] Cepheids in the Small Magellanic Cloud 
[SMC]: Kov\'acs 2000a [hereafter K00a]; preliminary study 
of the LMC beat Cepheids: Kov\'acs 2002). In these works 
we routinely assumed that the double-mode variables share 
the same common metallicity [Fe/H] as the rest of the 
given population of the cluster or galaxy. This is a crucial 
assumption, because the derived mass and luminosity 
(and therefore the distance) depend sensitively on [Fe/H]. 

The discovery of large number of first overtone Cepheids in 
the Magellanic Clouds (MCs) by the microlensing projects drew 
the attention to the evolutionary status of these objects. 
The problem with these stars is that they do not blue-loop 
into the instability strip as required by canonical Cepheid 
evolution scenarii, based mostly on longevity arguments at 
these looping periods of evolution. Recently, Cordier et al. 
(2003) concluded that, at least in the SMC, the only way to 
produce extended blue-loops at the expected luminosity level 
is to lower [Fe/H] by some $0.4$~dex.    

Beaulieu et al. (2001) (hereafter B01) discussed the systematic 
discrepancy between the theoretical and observed periods in 
the SMC beat Cepheids. They concluded that lowering the 
metallicity may ease this discrepancy but never eliminate 
it completely. They also mention that the distance modulus 
derived from beat Cepheids by K00a may not be correct, because 
of the use of the `optimum period fit' (as defined in that paper) 
instead of `exact period fit'.

In this paper we address the problem of the accurate period 
fit for the beat Cepheids in the MCs. We check the effect of 
a possible metallicity difference between the fundamental/first 
overtone (FU/FO) and first/second overtone (FO/SO) variables on 
the period fit and on the derived distance moduli.

%
%
\section{Method and database}

Our approach is basically identical with that of K00a. 
Briefly, we use the following set of relations to derive 
the mass $M$ and luminosity $L$, e.g., for FU/FO variables:
\begin{eqnarray}
P_0 & = & G_0(M,L,T_{\rm eff},Z) \\ 
P_1 & = & G_1(M,L,T_{\rm eff},Z) \\ 
\log T_{\rm eff} & = & H(V-I,\log g,Z) 
\end{eqnarray}
Here the functions $G_0$ and $G_1$ are given in a tabulated 
form as the outputs of our linear non-adiabatic (LNA), fully 
radiative pulsation code (Buchler 1990). Function $H$ is a 
simple linear relation when the appropriate parameter regime 
is fitted from the Castelli et al. (1997) stellar atmosphere 
models. The zero point is adjusted to the one derived from 
the infrared flux method by Blackwell \& Lynas-Gray (1994). 

Our LNA models contain 400 shells down to inner boundary, where 
$q_{\rm in}\equiv (R_{\rm in}/R_{\rm surf})(M_{\rm in}/M_{\rm surf})=0.05$. 
With this condition the innermost temperature was always greater 
that $2\times 10^6$~K but less than $\sim 10^7$~K. All models have 
$X=0.76$. With solar-type heavy element distribution, we computed 
7 model sequences with $Z=0.0003$, $0.001$, $0.002$, $0.003$, 
$0.004$, $0.008$, $0.01$. At each $Z$ we have the following 
equidistand model grid:
\begin{eqnarray} 
(T_{\rm eff}^{\rm min}, \Delta T_{\rm eff}, N_{T_{\rm eff}})
& = & (5000.0, 100.0, 21) \nonumber \\ 
(M^{\rm min},\Delta M, N_{\rm M}) 
& = & (2.0, 0.25, 17) \nonumber \\ 
(\log L^{\rm min}, \Delta \log L, N_{\rm L}) 
& = & (2.0, 0.1, 16) \nonumber  
\end{eqnarray} 
Altogether we computed nearly 40000 models for the 7 metallicity 
values. For accurate period fits we employed quadratic interpolation 
among the above models. Tests have shown that the interpolated 
periods often approximated the model periods with an accuracy of 
$\sim 10^{-4}$~d.   

The best matching $(M,L)$ parameters are searched by minimizing 
the following expression: 
\begin{eqnarray}
D & = & [\Delta \log P_0]^2 + [\Delta \log P_1]^2 \hskip 2mm ,
\end{eqnarray}
where the symbol $\Delta$ denotes the differences between the 
model and observed values. In cases when $D$ had two minima, 
for the FO/SO variables, the solution with $M<4.5$ was selected 
on evolutionary basis (by assuming that these stars are in the 
phase of second or third crossings). 

We used the OGLE Cepheid database for the periods and 
intensity-averaged colors as published by Udalski et al. (1999) 
and Soszy\'nski et al. (2000). We have 19 FU/FO and 57 FO/SO 
variables in the LMC. The corresponding figures for SMC are 23 
and 70, respectively. 

%
%
\section{Results}

First we test the probable cause of the discrepancy 
between the theoretical and observed periods found by B01. 
With the method described in Sect.~2, we compute average period 
ratio differences for the LMC FO/SO stars for model envelopes 
of various depth. As it is seen in Table~1, for shallow envelopes 
the period fit shows the same type of discrepancy as mentioned 
by B01. This discrepancy disappears for models with deep envelopes. 
From this test we conclude that the discrepancy found by B01 was 
caused most probably by the {\it shallowness} of their model 
envelopes. 
%
%
\begin{table}
\caption{\footnotesize 
Average observed minus computed period ratio as a function of 
the envelope depth. FO/SO stars in the LMC are tested at $Z=0.003$.}
\label{env}
\begin{flushleft}
\begin{tabular}{|r|r|r|r|}
\hline
$q_{\rm in}$ & 0.20 & 0.10 & 0.05 \\
\hline
$\Delta (P_2/P_1)$ & $ +0.00333 $ & $ +0.00053 $ & $ +0.00001 $ \\
\hline
\end{tabular}
\end{flushleft}
\end{table}

Next we check the dependence of the accuracy of the period fit 
on metallicity. Figs.~1 and 2 show this dependence for the LMC 
beat Cepheids. It is seen that if $Z$ is near to its canonical 
value, the FO/SO stars show systematic differences at higher 
period ratios, whereas the FU/FO stars, except for a few 
discrepant stars, show a good/exact fit. By lowering $Z$ brings 
nearly all FO/SO stars in exact fit but destroys the previous 
good fit for most of the FU/FO stars. This result implies that 
the two groups of beat Cepheids might have different metallicities 
and that chemical inhomogeneties might exist even within the 
same group. This latter statement is in agreement with the 
evolutionary results, because stars of higher luminosity 
(i.e., FU/FO variables) are able to blue-loop both at high and 
low metallicities.   

%
%
\begin{figure}[]
\resizebox{\hsize}{!}{\includegraphics[clip=true]{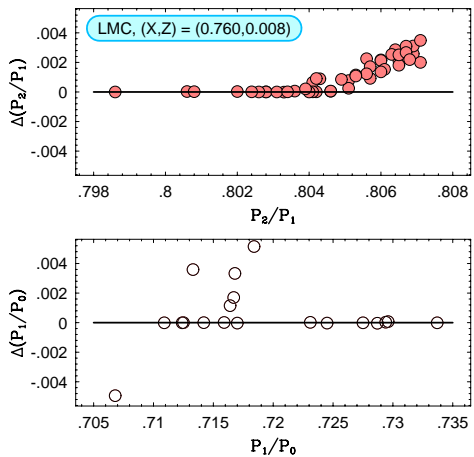}}
\caption{
\footnotesize
Observed minus computed period ratios vs. observed period ratios 
for the LMC beat Cepheids. The chemical composition used in the 
models is shown in the upper panel. FO/SO and FU/FO variables 
are plotted in the upper and lower panels, respectively.
}
\label{prdev008}
\end{figure}

%
%
\begin{figure}[]
\resizebox{\hsize}{!}{\includegraphics[clip=true]{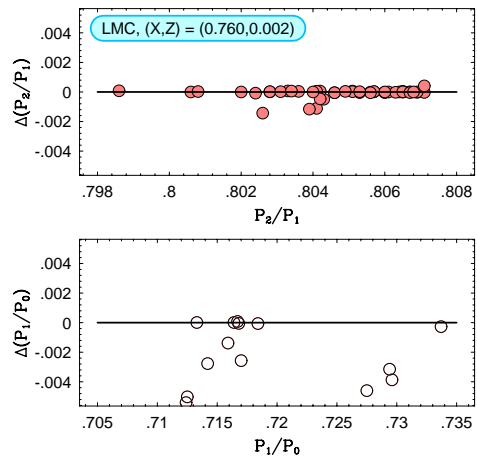}}
\caption{
\footnotesize
As in Fig.~1 but for lower $Z$.
}
\label{prdev002}
\end{figure}
%

%
%
\begin{figure}[]
\resizebox{\hsize}{!}{\includegraphics[clip=true]{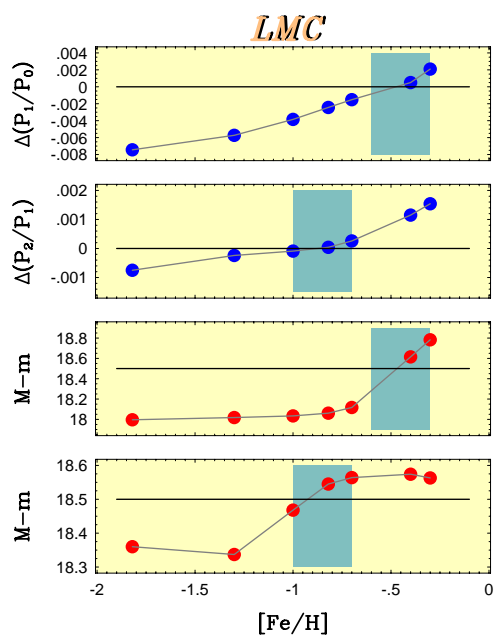}}
\caption{
\footnotesize
Metallicity dependence of the average period ratio difference 
(observed minus computed) and the distance modulus. Results for 
FU/FO and FO/SO variables are shown, from the bottom, in the 
second and fourth and in the first and third panels, respectively. 
Horizontal lines in the lower two panels indicate the canonical 
value of $18.5$~mag for the distance modulus. Shaded boxes mark 
the $\pm 0.15$~dex neighborhood of the exact overall period fit. 
}
\label{lmcdm}
\end{figure}

Assuming that variables within the same group have the same 
metallicity, we can study the effect of changing metallicity 
on the derived distance moduli. We see in Figs.~3 and 4 that 
the distance moduli derived from the FU/FO stars are much more 
sensitive to metallicity than those obtained from the FO/SO 
stars. These plots strongly suggest the existence of a systematic 
metallicity difference of $\sim0.4$~dex in both clouds between 
the two groups of beat Cepheids. Remarkably, at the metallicity 
regimes where the period fit is nearly exact, we obtain the same 
`canonical' distance moduli from both types of beat Cepheids.

%
%
\begin{figure}[]
\resizebox{\hsize}{!}{\includegraphics[clip=true]{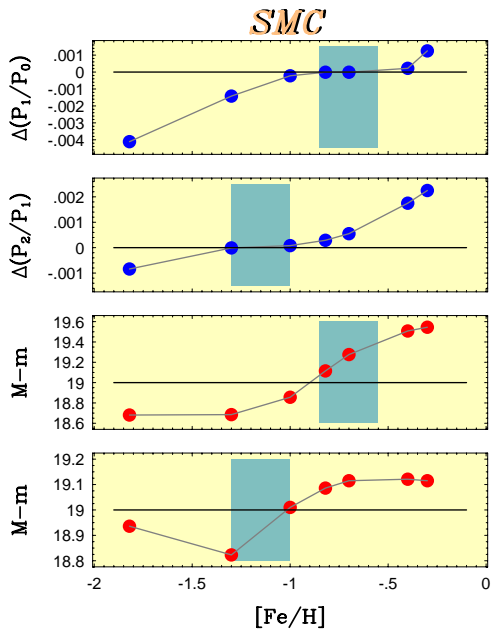}}
\caption{
\footnotesize
As in Fig.~3 but for the SMC. The horizontal lines in the lower 
two panels indicate the canonical value of $19.0$~mag for the 
distance modulus.
}
\label{smcdm}
\end{figure}

%
%
\section{Conclusions}

We have shown that exact match can be achieved between the 
linear pulsational model periods and the observed ones if the 
`canonical' metallicities of $-0.4$ and $-0.7$ are used for 
the FU/FO Cepheids in the LMC and SMC, respectively. On the 
other hand, FO/SO Cepheids can be fitted accurately only if 
their metallicities are lowered by some $0.4$~dex relative to 
the above values. The derived distance moduli from the FO/SO 
Cepheids are rather insensitive to these changes and they 
yield the same `canonical' values as the FU/FO Cepheids. 
These results are in agreement with those of Cordier et al. 
(2003) on the systematically lower metallicities of short-periodic 
Cepheids in the SMC. 

We also note that Moskalik \& Dziembowski (2005) analyzed two 
triple-mode Cepheids in the LMC. By searching for exact fit 
between the observed and model periods, they found that $Z$ 
could not be lower than $\sim0.004$, because otherwise not all 
the relevant modes would be excited. 

Here we only mention that for SC$3-360128$, $Z$ cannot be greater 
than $\sim0.004$ because then $T_{\rm eff}$ would be more than 
$400$K lower than estimated from the observed color index. It 
is also important to remark that when matching the observed 
periods with the ones derived from the LNA models, one needs 
to observe a `reasonable' error margin by which the periods 
can be fitted, because of the unknown difference between the 
LNA and nonlinear periods due to the lack of successful 
hydrodynamical modelling of FO/SO stars and, in particular, 
triple-mode ones.  

Direct metallicity measurements (Mottini 2005) will be very 
important in confirming the above results and in yielding more 
accurate observational constraints on future theoretical works.

\begin{acknowledgements}
This work has been supported by OTKA T-038437.  
\end{acknowledgements}

\bibliographystyle{aa}

\end{document}